\documentclass[prd,twocolumn,nofootinbib,showpacs,superscriptaddress]{revtex4}

\usepackage{graphicx}
\usepackage{amsmath,amssymb,amsfonts,revsymb}
\usepackage{epsfig}

\def\lsim{\mathrel{\rlap{\lower4pt\hbox{\hskip1pt$\sim$}}
    \raise1pt\hbox{$<$}}}
\def\gsim{\mathrel{\rlap{\lower4pt\hbox{\hskip1pt$\sim$}}
    \raise1pt\hbox{$>$}}}
\def\sqr#1#2{{\vcenter{\vbox{\hrule height.#2pt
         \hbox{\vrule width.#2pt height#1pt \kern#1pt
         \vrule width.#2pt}
         \hrule height.#2pt}}}}

\def\beq{\begin{equation}}
\def\eeq{\end{equation}}
\def\beqa{\begin{eqnarray}}
\def\eeqa{\end{eqnarray}}

\def\laq{\raise 0.4 ex \hbox{$<$}\kern -0.8 em\lower 0.62 ex\hbox{$\sim$}}
\def\gaq{\raise 0.4 ex \hbox{$>$}\kern -0.7 em\lower 0.62 ex\hbox{$\sim$}}

\begin{document}

\title{Generalized Chaplygin gas model, supernovae and cosmic topology}
\author{M.C. Bento}
\email{bento@sirius.ist.utl.pt}
\altaffiliation{Also at CFTP, Instituto Superior T\'ecnico, Av. Rovisco
  Pais, 1049-001 Lisboa}
\affiliation{%
Departamento de F\'{\i}sica, Instituto Superior T\'{e}cnico, Avenida
Rovisco Pais, 1049-001 Lisboa, Portugal}

\author{O. Bertolami}
\email{orfeu@cosmos.ist.utl.pt}
\affiliation{%
Departamento de F\'{\i}sica, Instituto Superior T\'{e}cnico,
Avenida Rovisco Pais, 1049-001 Lisboa, Portugal}

\author{M.J. Rebou\c{c}as}
\email{reboucas@cbpf.br}
\affiliation{%
Centro Brasileiro de Pesquisas F\'{\i}sicas  \\
Rua Dr.\ Xavier Sigaud 150 \\ 22290-180 Rio de Janeiro -- RJ, Brazil}

\author{P.T.  Silva}
\email{paptms@ist.utl.pt}
\affiliation{%
Departamento de F\'{\i}sica, Instituto Superior T\'{e}cnico,
Avenida Rovisco Pais, 1049-001 Lisboa, Portugal}

\date{\today}

\begin{abstract}

In this work we study to which extent
the knowledge of spatial topology may place constraints on the
parameters
 of the generalized Chaplygin gas (GCG) model for
unification of dark energy and dark matter.
By using both the Poincar\'e dodecahedral  and binary octahedral
spaces as the observable spatial topologies, we examine the current
type Ia supernovae (SNe Ia) constraints on the GCG model parameters.
We show that the knowledge of spatial topology does
provide  additional constraints on the $A_s$ parameter of the GCG
model but does not lift the degeneracy of the $\alpha$ parameter.
\end{abstract}



\maketitle

\section{Introduction}

Recently, the generalized Chaplygin gas (GCG) model
\cite{Kamenshchik:2001cp,Bilic:2001cg,Bento:2002ps} has attracted
considerable attention given its potential to account for
the observed accelerated expansion of the Universe \cite{observations},
and to describe in a simple scheme, both the negative pressure dark
energy component as well as the pressureless dark matter
component. In terms of the critical density, the contribution of each
component is about two thirds for dark energy and one third for dark
matter \cite{Bahcall}.

In the GCG proposal, the dark components are described through
a perfect fluid of density $ \rho_{ch}$ and pressure $p_{ch}$
with an exotic equation of state
\begin{align}
p_{ch} = - {A \over \rho_{ch}^\alpha}~, \label{rhoGCG}
\end{align}
where $A$ and $\alpha$ are positive constants.  For $\alpha=1$, the
equation of state is reduced to the Chaplygin gas scenario
\cite{Kamenshchik:2001cp}. The striking
feature of this model is that it allows for an unification of dark
energy and dark matter \cite{Bilic:2001cg,Bento:2002ps}.

The parameters of the GCG or indeed any dark energy model are known
to be affected by the spatial geometry the Universe.
Physicists describe the Universe as a manifold, which is characterized
by its geometry and its topology. Two fundamental questions regarding
the nature of the Universe concern the geometry and topology of the
$3$--dimensional space. Geometry is a local feature related with the
intrinsic curvature of the $3$--dimensional space and can be tested by studies
of the cosmic microwave background radiation (CMBR) such as the
Wilkinson Microwave Anisotropy Probe (WMAP).
Topology is a global property that
characterizes its shape and size. Geometry constrains but does not fix
the topology of the spatial sections. In a locally spatially homogeneous
and isotropic universe the topology of its spatial section dictates its
geometry.
Within the framework of the standard Friedmann--Lema\^{\i}tre--Robertson%
--Walker (FLRW) cosmology, the universe is modeled by a space-time manifold
$\mathcal{M}_4$ which is decomposed into $\mathcal{M}_4 = \mathbb{R}
\times M_3$ and endowed with a locally (spatially) homogeneous and
isotropic  metric
\begin{equation}
\label{RWmetric} ds^2 = -dt^2 + a^2 (t) \left [ d \chi^2 +
f^2(\chi) (d\theta^2 + \sin^2 \theta  d\phi^2) \right ] \;,
\end{equation}
where $f(\chi)=(\chi\,$, $\sin\chi$, or $\sinh\chi)$ depending on
the sign of the constant spatial curvature ($k=0,1,-1$).

The $3$--dimensional space where we live in is usually taken to be one
of the following simply-connected spaces: Euclidean $\mathbb{R}^3$,
spherical $\mathbb{S}^3$, or  hyperbolic space $\mathbb{H}^3$.
However, given that the connectedness of the spatial sections
$M_3$ has not been determined by cosmological observations, and
since geometry does not fix the topology, our $3$--dimensional space
may equally well be one of the possible multiply connected quotient
manifolds $M_3 = \widetilde{M}/\Gamma$, where $\Gamma$ is a fixed
point-free group of isometries of the covering space
$\widetilde{M}=(\mathbb{R}^{3},\mathbb{S}^{3}, \mathbb{H}^{3})$.

Thus, for instance, for the Euclidean geometry ($k=0$) besides
$\mathbb{R}^{3}$  there are 10 classes of topologically distinct
compact $3$--spaces consistent with this geometry, while for both the
spherical ($k=1$) and hyperbolic ($k=-1$) geometries there are an
infinite number of topologically inequivalent compact manifolds
with non-trivial topology that admit these geometries.

Recently, different strategies and methods to probe
a non-trivial topology of the spatial sections of the Universe have been
devised (see, e.g., the review articles Refs.~\cite{CosmTopReviews}
and also Refs.~\cite{CCmethods} for details on cosmic crystallographic
methods).
An immediate observational consequence of a detectable non-trivial
topology%
\footnote{The extent to which a non-trivial topology may have been
detected was discussed in Refs.~\cite{TopDetec}.}
of the $3$--dimensional space
$M_3$ is that the sky will exhibit multiple (topological)
images of either cosmic objects or specific spots on the CMBR.
The so-called ``circles-in-the-sky" method, for example, relies on
multiple  images of correlated circles in the CMBR maps~\cite{CSS1998}.
In a space with a detectable non-trivial topology, the sphere of last
scattering intersects some of its topological images along pairs of
circles of equal radii, centered at different points on the last
scattering sphere (LSS), with the same distribution 
of temperature fluctuations, $\delta T$. Since the mapping from the
last scattering surface to the night sky sphere preserves
circles~\cite{CGMR05}, these pairs of matching circles will be imprinted
on the CMBR anisotropy sky maps regardless of the background geometry
or detectable topology.
As a consequence, to observationally probe a non-trivial topology
one should scrutinize the full-sky CMBR maps in order to extract
the correlated circles, whose angular radii and relative position
of their centers can be used to determine the topology of the
Universe.
In this way, a non-trivial topology of the space section
of the Universe is observable, and can be probed for all locally
homogeneous and isotropic geometries. 

In this regard, in a recent work~\cite{RAMM} in the context of the
$\Lambda$CDM  model, the Poincar\'e  dodecahedral space was used
as the observable spatial topology of the Universe to reanalyze
the current type Ia supernovae (SNe Ia) constraints on the density
parameters associated with dark matter ($\Omega_m$) and dark energy
($\Omega_{\Lambda}$).  As a result, it has been shown that the
knowledge of the Poincar\'e dodecahedral space topology through the
``circles-in-the-sky" method gives rise to
stringent
constraints on the energy density parameters  allowed by the conventional
SNe Ia observations, reducing considerably the inherent degeneracies
of the current measurements.
Given this encouraging result it is natural to assess to what extent this
method can be useful for determining the parameters of
more complex dark energy models.
In this paper, we address these questions by focusing on the
constraints that cosmic topology
\footnote{In line with current literature,
by topology of the Universe we mean the topology of the spatial
section $M_3$.}
together with current SNe Ia data pose on the parameters of the
GCG model.
To this end, we use the Poincar\'e dodecahedral and
the binary octahedral spaces as the topologies of the spatial
sections of the Universe%
\footnote{These spatial topologies account for the low value
of the CMBR quadrupole and octopole moments
measured by the WMAP team, and fit
the temperature two-point correlation function, for
values of the total density within the reported range
~\cite{Poincare,Aurich1,Aurich2,WMAP-Spergel}.}
to reanalyze current constraints on the parameters
of the GCG model, as provided by the so-called \emph{gold}
sample of 157 SNe Ia~\cite{Riess:2004nr}.

\section{The Generalized Chaplygin Gas Model}

The integration of the energy conservation equation with the equation of
state (\ref{rhoGCG}), yields \cite{Bento:2002ps}
\begin{align}
\rho_{ch} = \rho_{ch0} \left[A_{s} + {(1-A_s) \over
a^{3(1+\alpha)}}\right]^{1/(1+\alpha)}~,
\end{align}
where $\rho_{ch0}$ is the present energy density of GCG and $A_s
\equiv A/\rho_{ch0}^{(1+\alpha)}$. One of the most striking
features of this expression is that the energy density,
 $\rho_{ch}$, interpolates between a dust dominated phase,
$\rho_{ch} \propto a^{-3}$, in the past and a de-Sitter phase,
$\rho_{ch} = -p_{ch}$, at late times. This property makes the GCG
model an interesting candidate for the unification of dark matter
and dark energy. Moreover,  one can see from the above equation
that $A_s$ must lie in the range $0\le A_s \le 1$: for $A_s =0$,
GCG behaves always as matter whereas for $A_s =1$, it behaves
always as a cosmological constant. We should point out, however,
that if one aims to unify dark matter and dark energy, one has to
exclude these two possibilities resulting in the range
$0< A_s < 1$. The value $\alpha = 0$ corresponds to the
$\Lambda$CDM model. Notice that in most phenomenological studies, the
range $0\le \alpha \le 1$ is considered, however it is shown that the
most recent supernova data favors $\alpha > 1$ values%
~\cite{Bertolami2004,Bento:2004ym}.

Friedmann's equation for a non-flat unified GCG model
is given  by \cite{Bento2003}
\begin{align}
\left({ {H} \over {H_0} }\right)^2 & =  \Omega_{r0}(1+z)^{4} +
\Omega_{b0}(1+z)^{3} + \Omega_{k}(1+z)^{2}\nonumber\\
& +  \Omega_{\mathrm{dark}} \left[A_{s} +
(1-A_s)(1+  z)^{3(1+\alpha)}\right]^{1/(1+\alpha)}~.
\label{GCG}
\end{align}
where $\Omega_{\mathrm{dark}}=1-\Omega_k-\Omega_{b0}-\Omega_{r0}$,
 $\Omega_k=1-\Omega_{\mathrm{tot}}$, $\Omega_{b0}=0.04$ and $\Omega_{r0} =
9.89 \times 10^{-5}$ are  the baryon and radiation energy density
contributions at present.
This model has been thoroughly scrutinized from the observational
point of view; indeed, its compatibility with the CMBR peak location
and amplitudes
\cite{Bento2003,Finelli}, with SNe Ia data
\cite{Bertolami2004,Bento:2004ym,Supern}, gravitational lensing statistics
\cite{Silva,Alcaniz} and gamma-ray bursts \cite{Bertolami2005}
has been extensively examined.

\section{Cosmic Topology Analysis}

The observed values of the power measured by WMAP of the CMBR quadrupole
($\ell=2$) and octopole ($\ell=3$) moments, and of the total
density $\Omega_{\mathrm{tot}}=1.02 \pm\, 0.02$ reported by WMAP
team~\cite{WMAP-Spergel}
have motivated the suggestion of the Poincar\'e dodecahedral space
topology as an explanation for the observed low power of
$\ell=2$ and $\ell=3$ multipoles~\cite{Poincare}.
Since then the dodecahedral space has been the scope of various
studies~\cite{Cornish,Roukema,Aurich1,Gundermann}, where
some important features have been  considered.
As a consequence, it turns out that a universe with the Poincar\'e
dodecahedral space section squares with WMAP data in that it accounts
for the suppression of power at large scales observed by WMAP, and
fits the WMAP temperature two-point correlation function
\cite{Aurich1,Aurich2},
retaining the standard FLRW description for local physics.

In a recent paper, Aurich \emph{et al.\/}~\cite{Aurich2}
have examined the behavior of both the CMBR angular power spectrum and the
two-point temperature correlation function for typical groups $\Gamma$
for which the spatial section $\mathbb{S}^3/\Gamma$ is (globally)
homogeneous. They have found that only three out of infinitely many
manifolds fit WMAP's low multipole ($\ell \leq 30$) power spectrum
the temperature correlations function, namely the Poincar\'e dodecahedron
$\mathcal{D}=\mathbb{S}^3/I^*\,$, $\mathcal{O}=\mathbb{S}^3/O^*\,$ and
$\mathcal{T}=\mathbb{S}^3/T^*\,$. Here $I^*$, $O^*$, and $T^*$ denotes,
respectively, the binary icosahedral group, the binary octahedral group,
and the binary tetrahedral group%
\footnote{A preliminary search failed to find the antipodal matched
circles in the WMAP sky maps predicted by the Poincar\'e
model~\cite{Cornish}. In a second search for these circles only a
non-conclusive indication for the correlated circles has been reported
for $\mathcal{D}$ and $\mathcal{T}$ spaces~\cite{Aurich3}.
Notice, however, that the Doppler and integrated Sachs-Wolfe contributions
may be strong enough to blur the circles, and thus the
correlated circles can be overlooked in the CMB sky maps search~\cite{Aurich1}.
In this way, the `absence of evidence may not be evidence of absence',
specially given that effects such as Sunyaev-Zeldovich, lensing and
the finite thickness of the LSS, as well as possible systematics in the
removal of the foregrounds, can further damage the topological circle
matching.} (for more details on the globally homogeneous spherical
manifold see the Appendix).

\begin{figure}[t]
\centerline{\psfig{figure=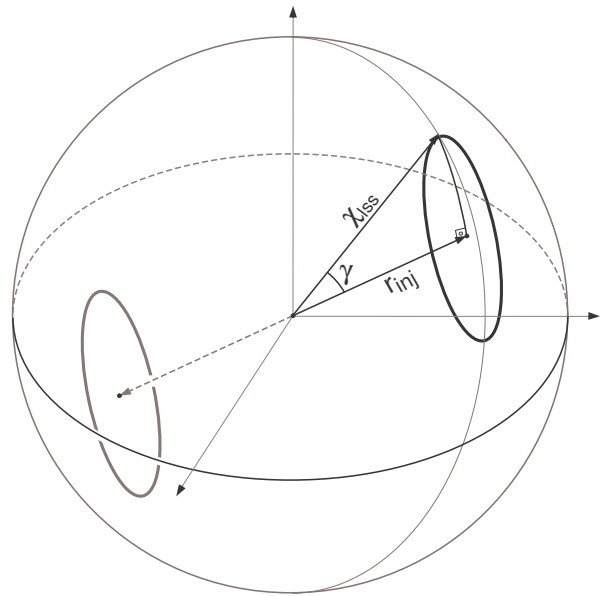,width=3.3truein,height=3.3truein,angle=0}
\hskip 0.1in} \caption{A schematic illustration of two antipodal
matching circles in the sphere of last scattering. The
relation between the angular radius $\gamma$ and the angular sides
$r_{inj}$ and $\chi^{}_{lss}$ is given by the following Napier's
rule for spherical triangles: $\cos \gamma = \tan
r_{inj}\, \cot \chi^{}_{lss}$~\cite{Coxeter}.}
\label{CinTheSky}
\end{figure}

Furthermore, the authors of  Ref.~ \cite{Aurich2} find that if
$\Omega_{\mathrm{tot}}$ is restricted to the interval [1.00, 1.04],
the space $\mathcal{T}$ is excluded since it requires a value of
$\Omega_{\mathrm{tot}}$ in the range [1.06, 1.07]. Thus, they
conclude that there remain only two globally homogeneous spherical
spaces that account for WMAP observed power spectrum, and fits the
WMAP temperature two-point correlation function, namely
$\mathcal{D}$ and $\mathcal{O}$.
In this paper, we shall restrict ourselves to the study of
the FLRW model with $\mathcal{D}$ and $\mathcal{O}$ sections.  We
begin by recalling that in the range of $\Omega_{\mathrm{tot}}$ where
they fit the WMAP data, these manifolds predict pairs of antipodal
matched circles in the LSS. Figure~\ref{CinTheSky} gives an
illustration of two of these antipodal circles.

\begin{figure}[t]
\centerline{\psfig{figure=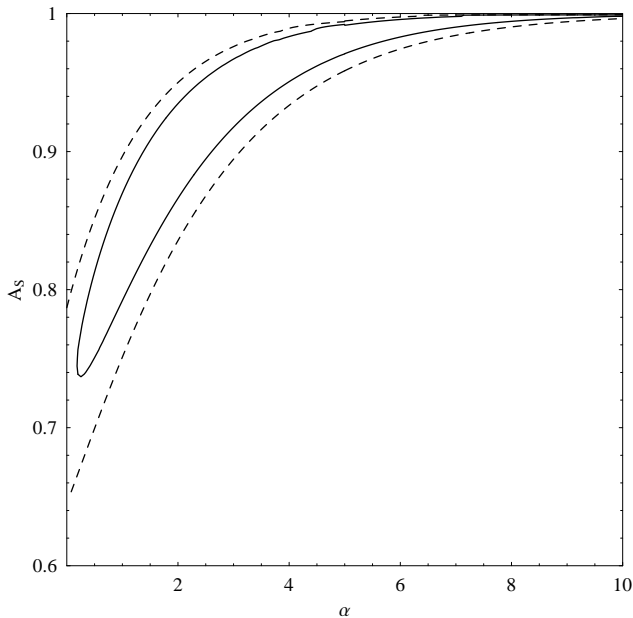,width=3.32truein,height=3.3truein,angle=0}
\hskip 0.1in} \caption{Confidence contours in the
$A_s-\alpha$ parameter space for the GCG model, using
the SNe Ia {\it gold} sample. The
solid and dashed lines represent
the $68\%$ and $95\%$ confidence regions, respectively.}
\label{justSN}
\end{figure}

\begin{figure*}[htb!]
\begin{center}
\includegraphics[height=7.5cm]{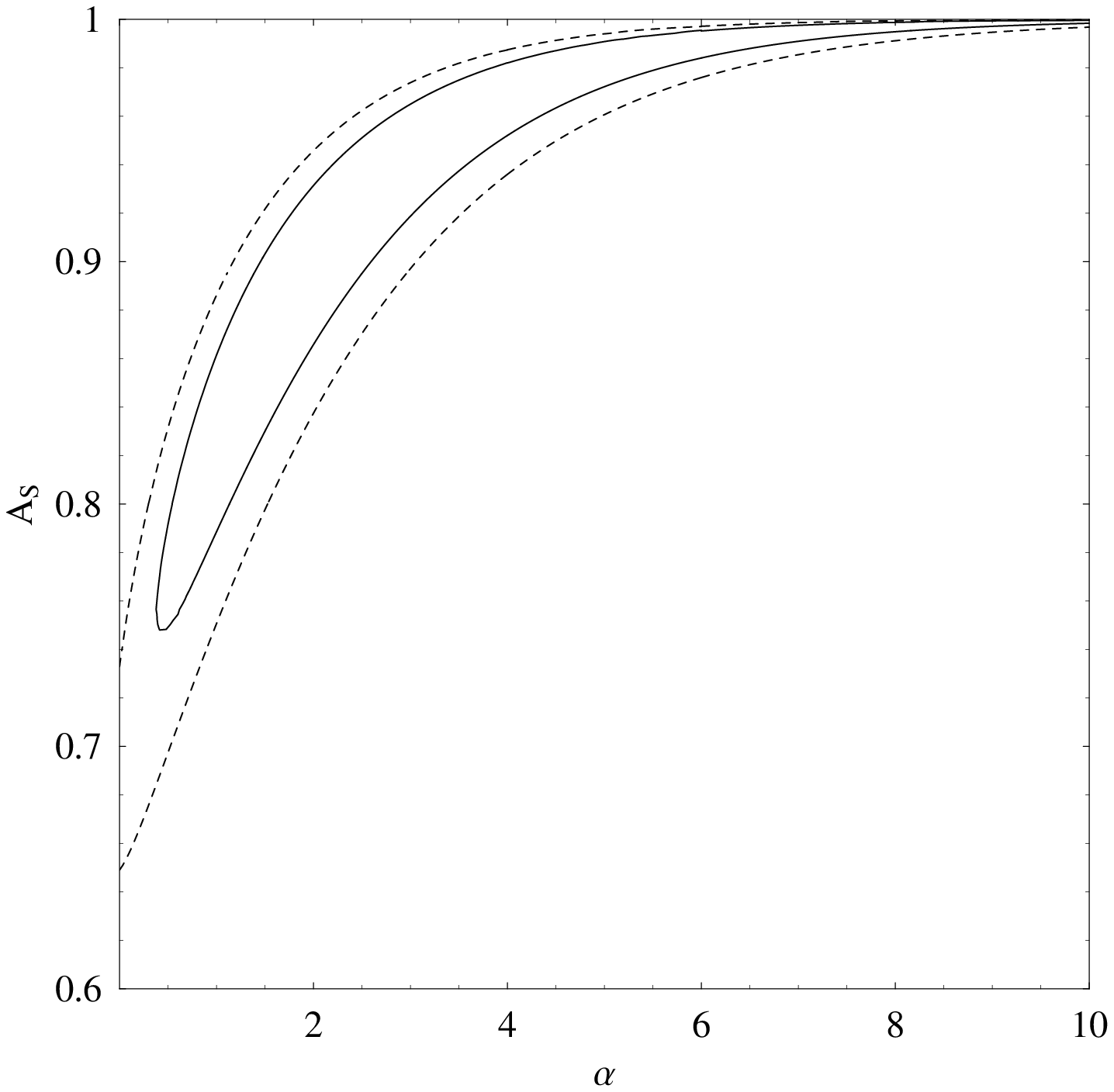}
\includegraphics[height=7.5cm]{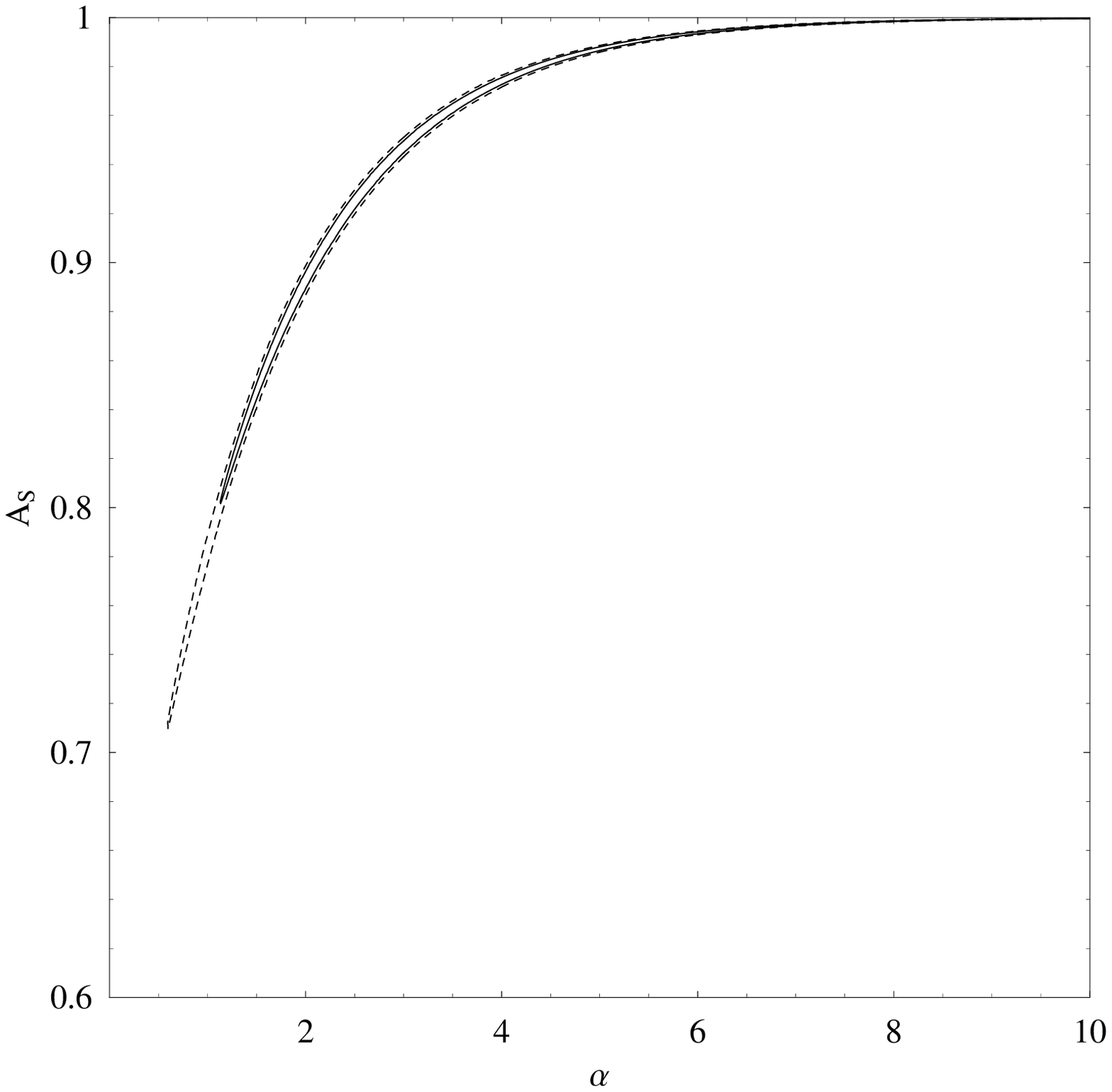}
\includegraphics[height=7.5cm]{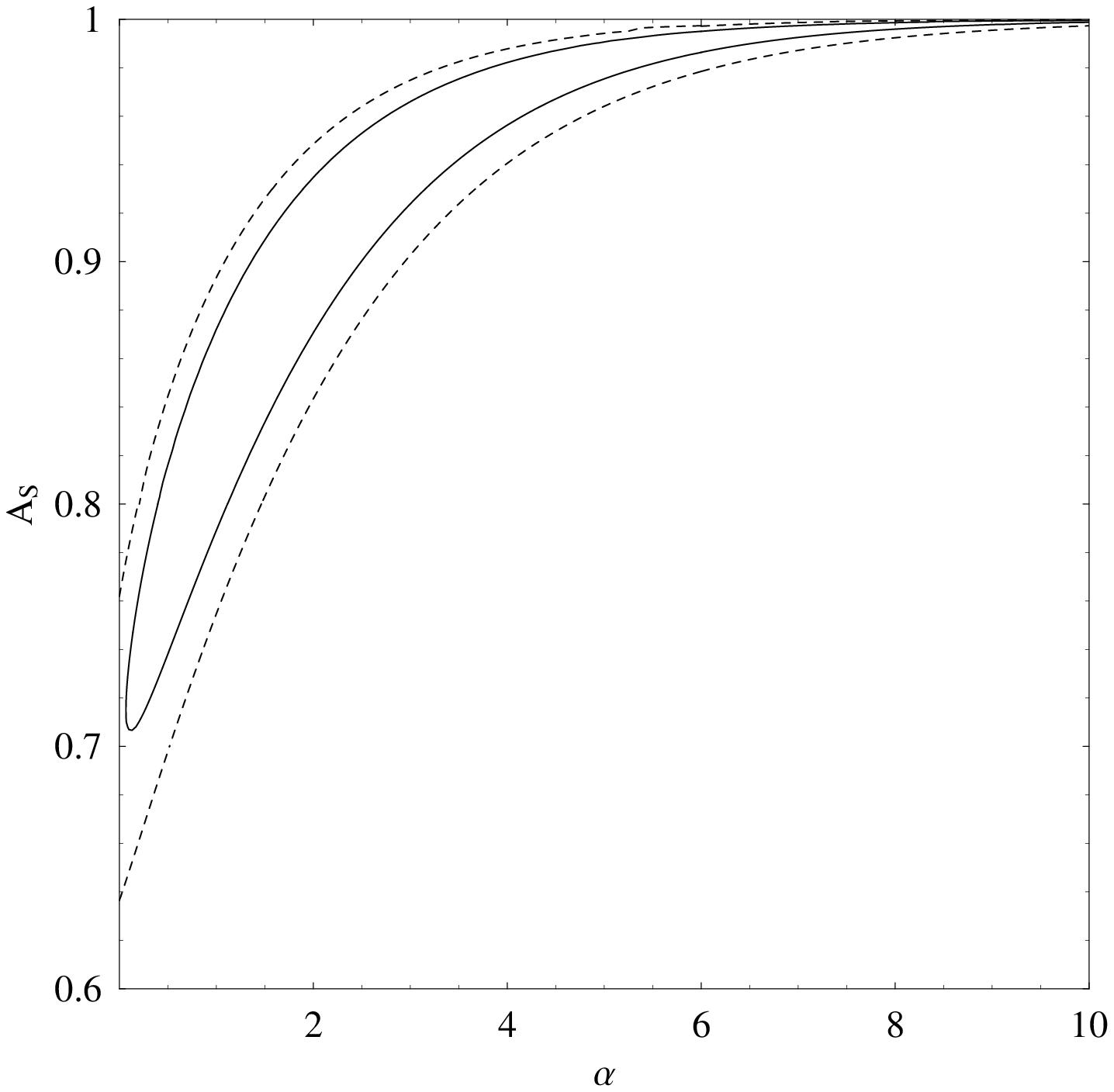}
\includegraphics[height=7.5cm]{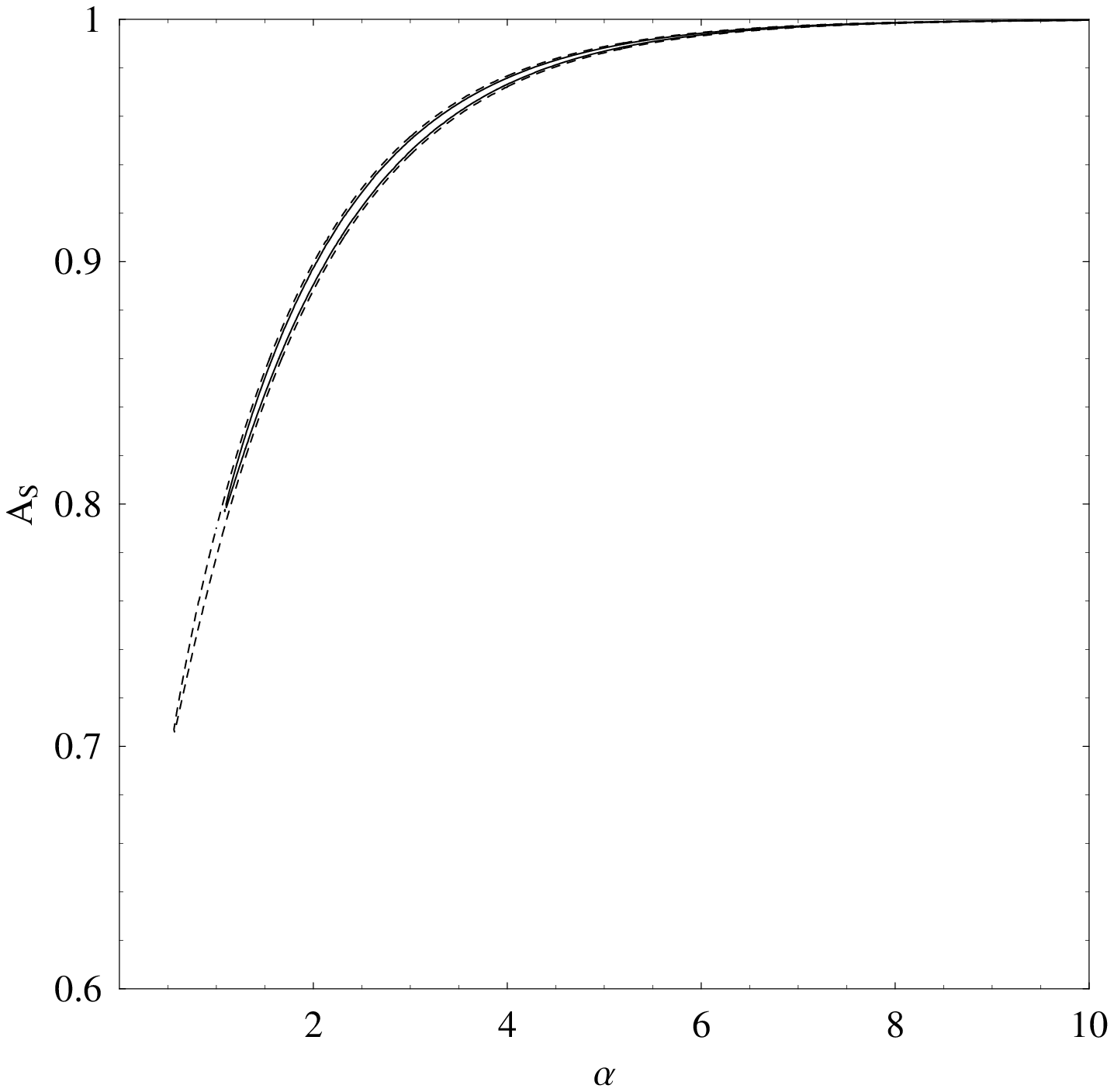}
\caption{\label{Dtopo} Confidence contours in the
$A_s-\alpha$ parameter space for the GCG model  using a joint
SNe Ia plus cosmic topology analysis.  The top and bottom
panels refer to the  $\mathcal{D}$ and $\mathcal{O}$ space
topology, respectively,  with angular radius
$\gamma=50^\circ\pm 6^\circ$ (left panel)
and $\gamma=11^\circ\pm 1^\circ$ (right panel). The
solid and dashed lines represent
the $68\%$ and $95\%$ confidence regions, respectively. Parameter
$\Omega_k$ is set at its best fit value in each case (see Table 1). }
\end{center}
\end{figure*}

The distance between the centers of each pair of circles
is twice the radius $r_{inj}$ of the smallest sphere inscribable in
the fundamental cells of these manifolds.
Now, a straightforward use of a Napier's rule on the right-angled
spherical triangle shown in Fig.~\ref{CinTheSky} gives  a relation
between the angular radius $\gamma$ and the angular sides $r_{inj}$ and
radius $\chi^{}_{lss}$ of the last scattering sphere, namely
\begin{equation}
\label{cosalpha}
\cos \gamma = \frac{\tan r_{inj}}{\tan \chi^{}_{lss} }\;,
\end{equation}
where $r_{inj}$ is a topological invariant, equal to
$\pi/10$ and $\pi/8$  for, respectively, $\mathcal{D}$ and $\mathcal{O}$.
This equation can be solved for $\chi^{}_{lss}$
to give
\begin{equation}
\label{Chigamma}
\chi^{}_{lss} = \tan^{-1} \left[\,\frac{\tan r_{inj}}{
\cos \gamma}\, \right] \;,
\end{equation}
where the distance $\chi^{}_{lss}$ to the origin \emph{in units of the
curvature radius},
$a_0=a(t_0)=(\,H_0\sqrt{|1-\Omega_{\mathrm{tot}}|}\,)^{-1}\,$,
is given by
\begin{equation}
\label{ChiLSS}
\chi^{}_{lss}= \frac{d^{}_{lss}}{a_0} = \sqrt{|\Omega_k|}
\int_1^{1+z_{lss}} \, \frac{H_0}{H(x)} \,\, dx \;,
\end{equation}
where $d^{}_{lss}$ is the radius of the LSS, $x=1+z$ is an integration
variable, $H$ is the Hubble parameter, $\Omega_k =
1-\Omega_{\mathrm{tot}}$, and
$z_{lss}=1089$~\cite{WMAP-Spergel}.
Eq.~(\ref{ChiLSS}) makes apparent that $\chi^{}_{lss}$ depends on
the cosmological scenario; moreover, Eq. (\ref{Chigamma}) with
$\chi^{}_{lss}$ given by Eq.~(\ref{ChiLSS}) together with
Eq.~(\ref{GCG}) allow us to find a
relation between the angular radius $\gamma$ and the cosmological
parameters of the model. Thus, they can be used to set bounds
(confidence regions) on these parameters. To quantify this we proceed
in the following way.
Firstly, for a comparative study we consider a typical angular radius
$\gamma = 50^\circ$ estimated in Ref.~\cite{Aurich1} for the Poincar\'e
dodecahedral space. Secondly, we  note that measurements of the radius
$\gamma$ unavoidably involve observational uncertainties, and therefore,
in order to set constraints on the density parameters from the detection
of cosmic topology, one should take such uncertainties into account.
In order to obtain conservative results we consider
$\delta \gamma \simeq 6^\circ$,
which is the scale below which the circles are blurred in the dodecahedron
case~\cite{Aurich1}.  We also analyze the case
$\gamma = 11^\circ \pm 1^\circ$, as suggested in Ref.~\cite{Roukema}.

\begin{figure*}[htb!]
\begin{center}
\includegraphics[height=6.5cm]{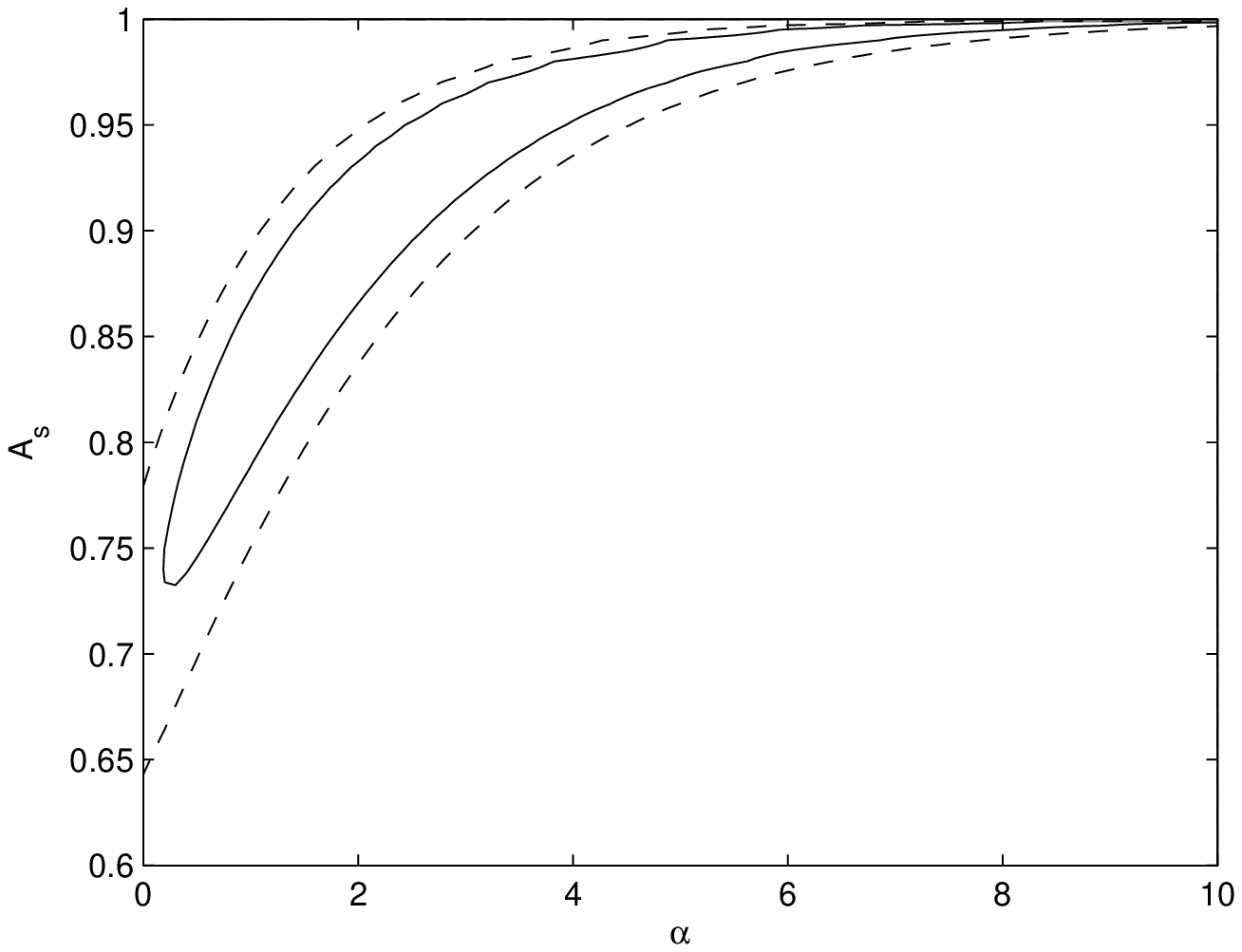}
\includegraphics[height=6.5cm]{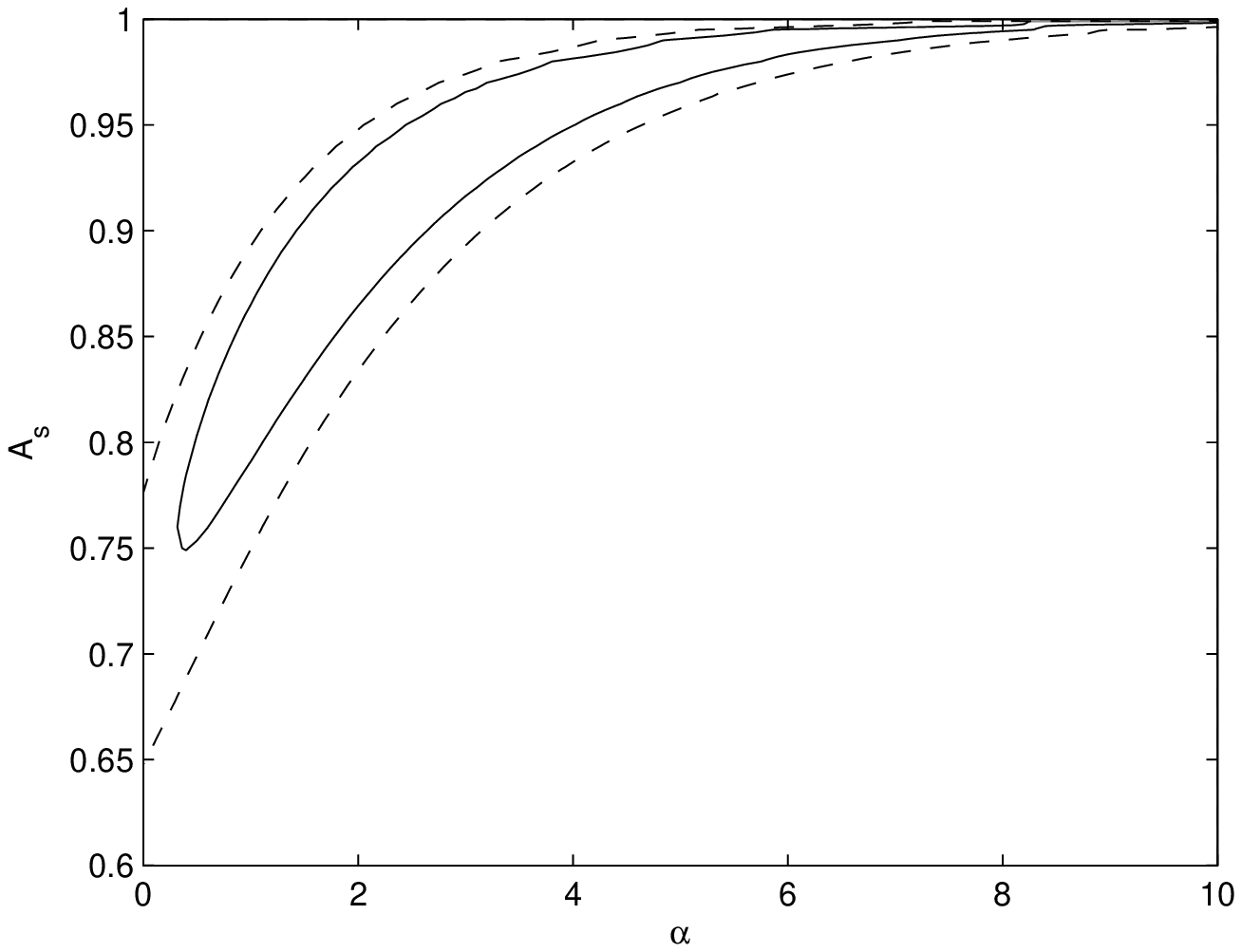}
\includegraphics[height=6.5cm]{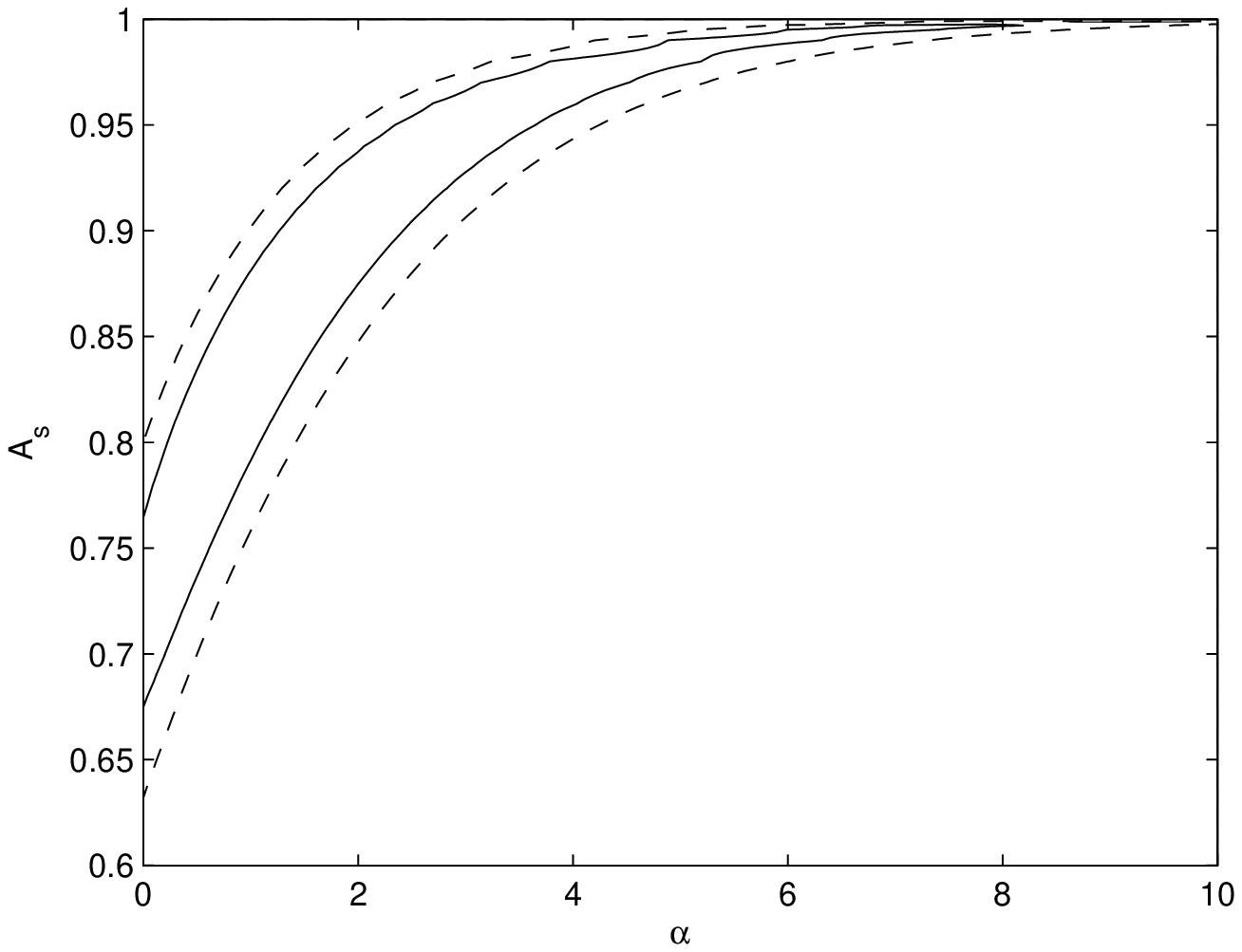}
\includegraphics[height=6.5cm]{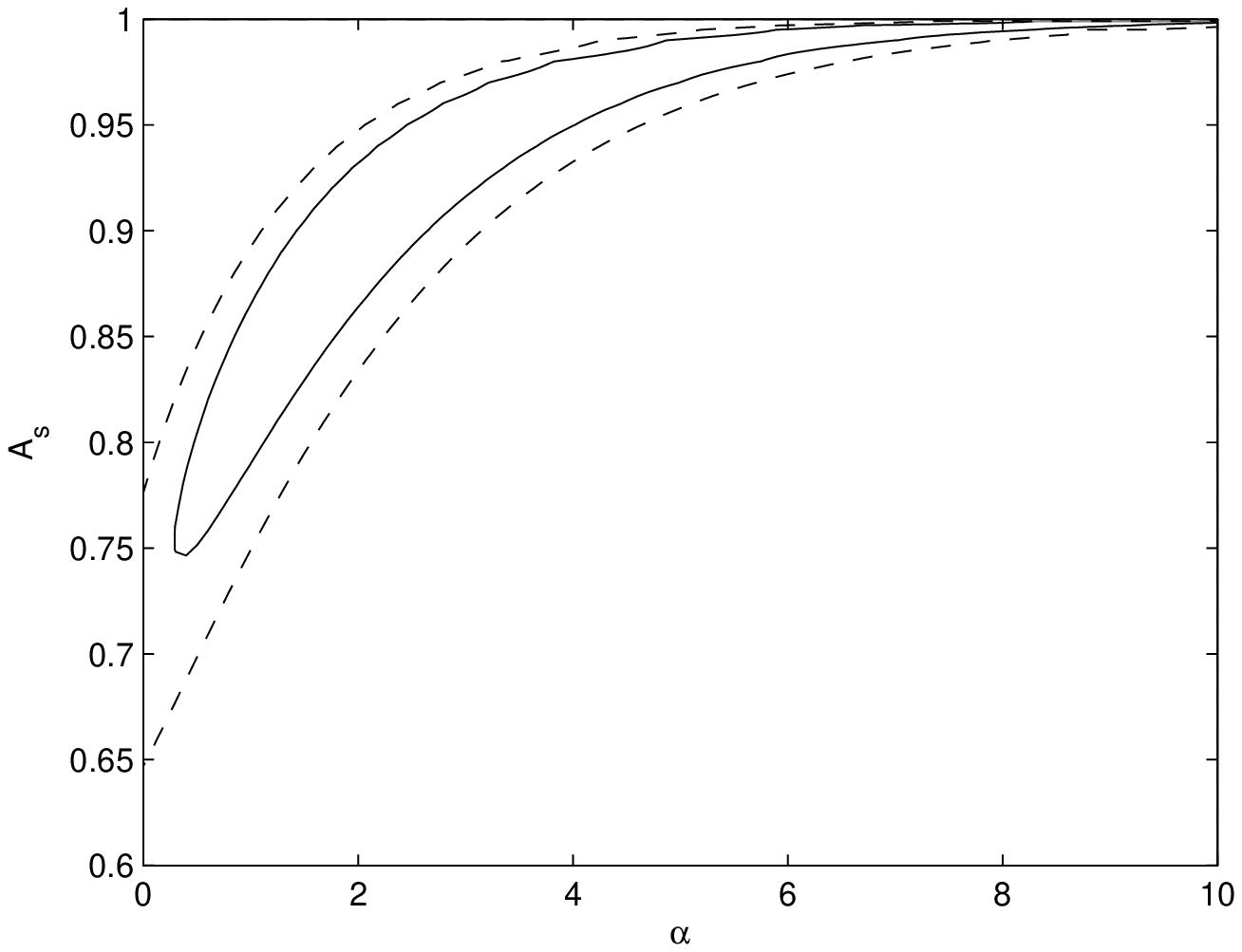}
\caption{\label{Otopo} As for Fig. \ref{Dtopo} but parameter $\Omega_k$
has been marginalized over.}
\end{center}
\end{figure*}

\section{Observational constraints from supernovae data and cosmic topology}

The observations of supernovae measure essentially the
apparent magnitude $m$, which is related to the luminosity distance $d_L$ by
\begin{align}
m(z) = {\cal M} + 5 \log_{10} D_L(z) ~,
\end{align}
where
\begin{align}
D_L(z) \equiv {1 \over{c}} d_L(z)~, \label{DL}
\end{align}
is the dimensionless luminosity distance and
\begin{align}
d_L(z)=(1 + z) d_M(z)~,
\label{dL}
\end{align}
with $d_M(z)$ being the comoving distance, given by
\begin{align}
d_M(z)={c \over \sqrt{|\Omega_k|}}S_k \left(\sqrt{|\Omega_k|}~H_0 \int_0^z
{{1}\over{H(z')}} dz'\right)~, \label{dm}
\end{align}
where $S_k(x) = \sin x$ if $\Omega_k < 0$, $S_k(x) = \sinh x$ if
$\Omega_k > 0$ and $S_k(x)= x$ if  $\Omega_k =0$. Furthermore,
\begin{align}
{\cal M} = M + 5 \log_{10}
\left({{c/H_0}\over{1~\mbox{Mpc}}}\right) + 25~,
\end{align}
where $M$ is the absolute magnitude which is believed to be constant for
all SNe Ia.

For our analysis, we consider the set of SNe Ia data recently compiled
by Riess {\it et al.} \cite{Riess:2004nr} known as the {\it gold}
sample. This set contains 143 points from previously published data
that were taken from the 230 Tonry {\it et al.} \cite{Tonry:2003zg}
data along with the 23 points from Barris {\it et al.}
\cite{Barris:2003dq}. In order to increase the reliability of the
sample, various points where the classification of the supernovae was
unclear or the photometry was incomplete were discarded. The {\it
gold} sample contains also 14 points recently discovered using the
Hubble Space Telescope consisting altogether of 157 points
\cite{Riess:2004nr}.  The data points in these samples are given in
terms of the distance modulus

\begin{align}
\mu_{\rm obs}(z) \equiv m(z) - M_{\rm obs}(z)~,
\end{align}
and the  $\chi^2$ is calculated from

\begin{align}
\chi^2 = \sum_{i=1}^n \left[ {{\mu_{\rm obs}(z_i) - {\cal M}' - 5
\log_{10}D_{L \rm th}(z_i; \alpha, A_s)}\over{\sigma_{\mu_{\rm
obs}}(z_i)}} \right]^2~,
\label{chisq2}
\end{align}
where ${\cal M}' = {\cal M} - M_{\rm obs} $ is a free parameter
and $D_{L \rm th}(z;\alpha, A_s)$ is the theoretical prediction
for the dimensionless luminosity distance of a supernova at a
particular distance, for the GCG model with parameters
$\alpha, ~A_s$, which  can be computed using the Friedmann
expansion rate (see below) combined with Eqs.~(\ref{DL})--(\ref{dm}).
The errors $\sigma_{\mu_{\rm obs}}(z)$  take into
account the effects of peculiar motions.
\begin{table}[t!]
\begin{tabular}{c c c c c c c}
 \hline \hline
SNe Ia & Topology & \hspace*{4mm} $\gamma$ \hspace*{4mm}  &  $A_s$ & $\alpha$
                                                & $\Omega_{k}$ & $\chi^2$ \\
\hline
 {\it Gold} sample  &  $-$ & $-$ & $0.95$ & $3.07$ & $0.00$ & $174.2$ \\
\hline
{\it Gold} sample & $\mathcal{D}$ & $50^\circ$ & $0.93$ & $2.58$
& $-0.031$ & $174.3$ \\
&  & $11^\circ$ & $0.94$ & $2.83$ & $-0.014$ & $174.3$ \\
\hline
{\it Gold} sample & $\mathcal{O}$ & $50^\circ$ & $0.89$ & $1.74$
& $-0.040$& $174.3$ \\
& &  $11^\circ$  & $0.94$ & $2.70$ & $-0.023$ & $174.3$ \\
 \hline\hline
\end{tabular}
\caption{Best fit parameters for the GCG
model, for a SNe Ia and joint SNe Ia plus cosmic topology analysis,
namely  the space
topologies $\mathcal{D}$  and $\mathcal{O}$.}
\label{table:best}
\end{table}
We have performed a best fit analysis with the minimization of
the $\chi^2$, Eq.~(\ref{chisq2}), with respect to $\Omega_k$ and
the GCG model parameters, using a MINUIT \cite{Minuit} based code.

The $\mathcal{D}$ or the $\mathcal{O}$ spatial topology is added
to the conventional SNe Ia data analysis as a Gaussian
prior on the value of $\chi^{}_{lss}$, which can be easily obtained
from an elementary combination of Eqs.~(\ref{Chigamma})--(\ref{ChiLSS})
taking into account the ratio $H_0/H$ for the GCG model.
In other words, the contribution of the topology to $\chi^2$ is a term
of the form $\chi^2_{\mathrm{topology}} = (\chi^{\mathrm{Obs}}_{lss} -
\chi^{\mathrm{Th}}_{lss})^2 / (\delta \chi_{lss})^2$, where
$\chi^{\mathrm{Th}}_{lss}$ is given by Eq.~(\ref{Chigamma}) and
$\delta \chi_{lss}$ is the uncertainty considered in the
``circles-in-the-sky" method.

To find the desired confidence regions, we must eliminate the
dependence of the $\chi^2$ function on the nuisance parameter
$\cal{M}'$, and the curvature $\Omega_k$. We first consider
the elimination of the nuisance parameter $\cal M'$.
One way to approach this problem consists in minimizing the
$\chi^2$ function, and fixing the value of $\cal M'$ to the
value corresponding to the minimum of the $\chi^2$ function.
An alternative method consists in marginalizing the likelihood
function associated with the $\chi^2$ function over the unwanted
parameter, using some probabilistic prior $\pi({\cal M'})$. Using
this method, one finds a modified $\widetilde{\chi}^2$ function given by
\begin{align}
\widetilde{\chi}^2(\theta) =  -2\ln \int\left[ \exp\left
(-{\chi^{2}(\theta,M') \over 2}\right)
\pi ( {\cal M'} ) d{\cal M'} \right],
\end{align}
where $\theta$ stands for the other cosmological parameters.

We marginalize over the nuisance parameter for all cases. We have placed
no prior on $\cal M'$, that is, we considered that all values are
equally likely.

As for the curvature, we have used both methods (cf.\ Figures \ref{Dtopo}
and \ref{Otopo}). When marginalizing over $\Omega_k$ we used the uniform
prior that $\Omega_k \in [-0.04,0.0[\,$, obtained from WMAP's reported
range for the total energy density $\Omega_{\mathrm{tot}}$~\cite{WMAP-Spergel}.
By using both methods we can have an idea of the sensitivity of the
test regarding the curvature parameter. Given that
the results are very different
for each method (see Figures \ref{Dtopo} and \ref{Otopo}), we conclude
that the test is quite sensitive to the parameter we are marginalizing over.

In Table 1, we summarize the results of our best fit analysis.
Figure~\ref{justSN}  shows the results of the SNe Ia analysis
alone (no cosmic topology prior).
The full and dashed curves represent, respectively, the
$68.3\%$ and $95.4\%$ confidence regions in the $\alpha-A_s$
parametric plane. In Figure~\ref{Dtopo}, we show the results of our
joint SNe Ia
plus cosmic topology analysis for the case of the  $\mathcal{D}$ (top
panel) and  $\mathcal{O}$ (bottom panel) space topologies,
with angular radius $\gamma=50^\circ\pm 6^\circ$ (left panel) and
$\gamma=11^\circ\pm 1^\circ$ (right panel).

Our results indicate that
the combination of SNe Ia data with the detection of either
$\mathcal{D}$ or $\mathcal{O}$ spatial topology through the
so-called ``circles-in-the-sky" method yield some
constraints on $A_s$, which become
more important for small values of the angular radius. These
constraints are tighter
for the $\mathcal{D}$ space topology than for  the $\mathcal{O}$ space
topology. Indeed, in the former case, we find that $0.75 \lsim A_s \lsim 1$
while in the latter $0.7 \lsim A_s \lsim 1$, at $68.3\%$ C.L. and for
$\gamma = 50^\circ$. For $\gamma = 11^\circ$, bounds
are tighter, $0.8 \lsim A_s \lsim 1$, for both spatial topologies.
These limits are consistent with bounds that can be derived
(for the best fit value of $\Omega_k$) by superimposing the contour
curves $\chi^{}_{lss}(A_s, \alpha)= r_{inj}$ for $\mathcal{D}$
and $\mathcal{O}$ on the region of the $\alpha$~--~$A_s$ plane
allowed by the SNe Ia data~\cite{MMR2005}.
Also notice that the $\mathcal{D}$ space topology is slightly less
curved than the $\mathcal{O}$ space topology.

As for the $\alpha$ parameter, we find that it is highly degenerated
and, likewise other phenomenological tests, the ``circles-in-the-sky"
method does not lift this redundancy significantly for the spatial
topologies we have analyzed. Actually, so far it has been only through
studies of structure formation that a significant dependency on the
$\alpha$ parameter has been found (see \cite{Bento:2004sf} and
references therein).  In any case, consistently with the most recent
supernova data analysis for the GCG
model~\cite{Bertolami2004,Bento:2004ym}, we find that the most likely
values for $\alpha$ are greater than one. As for the consistency of
our analysis with the one for the $\Lambda$CDM model of
Ref. \cite{RAMM} we have verified that our results match the ones of
that study in the limit $\alpha=0$ leaving $\Omega_k$ free.

Finally, we would like to remark on three
important features of our results. First, that the best-fit values are just
weakly dependent on  angular radius $\gamma$ of the circle.
Second, that the uncertainty on the value of the radius $\gamma$
alters predominantly the area corresponding to the confidence regions,
without having a significant effect on the best-fit values.
Third, there is a topological degeneracy in that the same best fits and
confidence regions found for e.g. the $\mathcal{D}$ topology
arise from either the $\mathcal{Z}_{10}=\mathbb{S}^3/Z_{10}$ or
the $\mathcal{D}_5=\mathbb{S}^3/D^*_5$ globally
homogeneous spherical spatial topologies.
Similarly, $\mathcal{O}$,  $\mathcal{Z}_8=\mathbb{S}^3/Z_{8}$ and
$\mathcal{D}_4=\mathbb{S}^3/D^*_4$ give
rise to identical bounds on the GCG parameters. Here $Z_n$ and
$D^*_m$ denotes, respectively, the cyclic and dihedral groups.

\section{Conclusions and outlook}

The so-called ``circles-in-the-sky" method makes apparent that a
non-trivial detectable topology of the spatial section of the Universe
can be probed for any locally homogeneous and isotropic universe, with
no assumption on the cosmological density parameters.  In this paper
we have shown that the knowledge of $\mathcal{D}$ and $\mathcal{O}$
spatial topologies does provide some additional constraints on the
$A_s$ parameter of the GCG model, even though it does not help in
lifting the degeneracy on the $\alpha$ parameter.

In any case, our results indicate that the introduction of topological
considerations into the analysis of the large scale structure of the
Universe is an interesting complementary strategy to constrain and
eventually characterize the nature of dark energy and dark matter. In
the particular case of the GCG, the complexity of the model does not
allow for obtaining striking constraints on its parameters as is
the case for the $\Lambda$CDM model.  Finally, the question arises
whether topology may play a significant role for other dark energy and
modified gravity models, an issue we plan to analyze in a future
publication.

\begin{acknowledgments}

MJR thanks CNPq for the grants under which this work was carried out.
The work of M.C.B. and O.B. was partially supported by Funda\c c\~ao para a
Ci\^encia e a Tecnologia (FCT, Portugal) under the grant POCI/FIS/56093/2004.
\end{acknowledgments}

\appendix*
\section{}

Within the framework of FLRW cosmology, the Universe is
modeled by a $4$-manifold $\mathcal{M}_4$ which is decomposed into
$\mathcal{M}_4 = \mathbb{R} \times M_3$, and is endowed with a locally
homogeneous and isotropic Robertson--Walker  metric, Eq.~(\ref{RWmetric}).
The spatial section $M_3$ is usually taken to be one of the following
simply-connected spaces: Euclidean $\mathbb{R}^{3}$ ($k=0$),
spherical $\mathbb{S}^{3}$ ($k=1$), or hyperbolic $\mathbb{H}^{3}$
($k=-1$) spaces. However,
$M_3$ may equally well be any one of the possible
quotient (multiply-connected) manifolds $\mathbb{R}^3/\Gamma$,
$\mathbb{S}^3/\Gamma$, and $\mathbb{H}^3/\Gamma$, where $\Gamma$ is
a fixed-point free discrete group of isometries of the
covering space $\mathbb{R}^3$, $\mathbb{S}^3$ and $\mathbb{H}^3$.

The action of $\Gamma$ tiles the corresponding covering space
$\mathbb{R}^3$, $\mathbb{S}^3$ and $\mathbb{H}^3$, into
identical cells or domains which are copies of the so-called fundamental
polyhedron (FP). A FP plus the face identifications given by the group
$\Gamma$ is a faithful representation of the quotient manifold $M_3$.
An example of quotient manifold in three dimensions is the flat $3$--torus
$T^3=\mathbb{S}^1 \times \mathbb{S}^1 \times \mathbb{S}^1=\mathbb{R}^3/\Gamma$.
The  covering space clearly is $\mathbb{R}^3$, and the FP is a cube with
opposite faces identified after a translation. This FP tiles the covering
space $\mathbb{R}^3$. The group $\Gamma=\mathbb{Z} \times \mathbb{Z} \times
\mathbb{Z}$ consists of discrete translations associated with the face
identifications.

An important topological feature of the spherical and hyperbolic
$3$--manifolds $M_3$ is the so-called injectivity radius $r_{inj}$,
which corresponds to the radius of the smallest sphere that can be inscribed
in $M_3$, which can be formally defined in terms of the length of the
smallest closed geodesics $\ell_M\,$ by $r_{inj} = \ell_M/2$.

\begin{table}[!hbt]
\begin{center}
\begin{tabular}{|cccc|}
\hline
\ Name  & Covering Group $\Gamma$ & \ Order of $\Gamma$ \ & \ $r_{inj}$ \  \\
\hline \hline
$\mathcal{Z}_n$ &   Cyclic              $Z_n$   & $n$  & $\pi/n$           \\
$\mathcal{D}_m$ &   Binary dihedral     $D^*_m$ & $4m$ & $\pi / 2m $      \\
$\mathcal{T}$   & \ \ Binary tetrahedral  $T^*$    & 24   & $ \pi/6$          \\
$\mathcal{O}$   & \ \ Binary octahedral   $O^*$    & 48   & $\pi/8$          \\
$\mathcal{D}$   & \ \ Binary icosahedral  $I^*$    & 120  & $\pi/10$         \\
\hline
\end{tabular}
\end{center}
\caption{The globally homogeneous spherical manifolds,
$M_3=\mathbb{S}^3/\Gamma$, along with their  covering
groups, $\Gamma$, the order of $\Gamma$ and  the injectivity
radius $r_{inj}$.
The cyclic and binary dihedral cases  constitute families of
manifolds, whose members are given by the different values of the
integers $n$ and $m$.}
\label{SingleAction}
\end{table}

In this work we focus our attention in globally homogeneous spherical
manifolds.
The multiply connected spherical $3$-manifolds are of the form
$M_3=\mathbb{S}^3/\Gamma$, where $\Gamma$ is a finite fixed-point
free subgroup of $SO(4)$.  The order of $\Gamma$ gives the number of
fundamental polyhedra  needed to fulfill the whole covering space
$\mathbb{S}^3$. These manifolds were originally classified
in Ref.~\cite{ThrelfallSeifert} (for a description in the context
of cosmic topology see the pioneering article by Ellis \cite{Ellis71}).
Such a classification consists essentially in the enumeration of all finite
groups  $\Gamma \subset SO(4)$, and then in grouping the possible manifolds
in classes. In a recent paper~\cite{GLLUW} the classification has been
recast in terms of single action, double action, and linked action
manifolds. Single action manifolds are globally homogeneous, and then
satisfy a topological principle of (global) homogeneity, in the sense
that all points in $M$ share the same topological properties.
In Table~\ref{SingleAction} we list the single action manifolds
together with the symbol we use to refer to them, the covering groups
$\Gamma$ and their order as well as the corresponding injectivity radius
$r_{inj}$.
We point out that the binary icosahedral group $I^{\ast}$ gives rise
to the known Poincar\'e dodecahedral space $\mathcal{D}$, whose
FP is a regular spherical dodecahedron,
$120$ of which tile the $3$-sphere into identical cells which
are copies of the FP. The FP of the $\mathcal{O}$ space is the
truncated cube, $48$ of which tile the sphere $\mathbb{S}^3$.

An important point concerning the spherical manifolds is that the
injectivity radius $r_{inj}$ expressed in \emph{units of the
curvature radius} is a constant (topological invariant) for a
given manifold $M$.

\end{document}